# Epitaxial growth of topological insulator $\beta$-Ag$_2$Te thin films

Ayuki Takegawa[1], Kouya Imoto[2], Minoru Kawamura[3], Moeta Tsukamoto[1], Ryutaro Yoshimi[1*]

[1] Department of Advanced Materials Science, The University of Tokyo, Kashiwa, Chiba, 227-8561, Japan

[2] Department of Applied Physics, The University of Tokyo, Bunkyo-ku, Tokyo 113-8656, Japan

[3] RIKEN Center for Emergent Matter Science (CEMS), Wako Saitama 351-0198, Japan.

* Corresponding author: r-yoshimi@edu.k.u-tokyo.ac.jp

# ABSTRACT

We report epitaxial growth of $\beta$-Ag$_2$Te thin films by molecular beam epitaxy. $\beta$-Ag$_2$Te, recently identified as a topological insulator, was grown by depositing Ag on InP substrate at room temperature followed by Te supply at elevated temperature. X-ray diffraction measurements and transmission electron microscopy analyses confirmed the (002) crystal orientation and the epitaxial atomic arrangement of $\beta$-Ag$_2$Te thin films. Electrical transport measurements revealed that the $\beta$-Ag$_2$Te thin film exhibits two-dimensional metallic conduction while the bulk remains insulating. The epitaxial $\beta$-Ag$_2$Te thin films obtained here provide a viable platform for investigating emergent phenomena arising from surface Dirac states and for designing heterojunction-based device structures.

**MAIN TEXT**

Three-dimensional topological insulators (TIs), which possess insulating bulk states and conducting surface states, have attracted significant interest in condensed matter physics since their discovery [1-4]. The surface states are characterized by exotic electronic properties such as linear band dispersion with spin-momentum locking, a nontrivial Berry phase, and robustness against nonmagnetic disorder due to topological protection. When breaking time reversal symmetry at the surface states, a variety of emergent phenomena arise, including the quantum anomalous Hall effect [5-7], unidirectional magnetoresistance [8], topological magnetoelectric effect [9,10] and current-driven magnetization switching [11,12]. One effective approach to realize these phenomena is to form heterojunctions composed of ferromagnetic insulators or metals (FMIs/FMMs) and TI epitaxial films. Such heterostructures enable simultaneous breaking of time-reversal and spatial-inversion symmetries, while maintaining high interfacial crystallinity. Indeed, a wide variety of FMIs/FMMs, such as (Zn,Cr)Te, $Cr_2Ge_2Te_6$, EuS or $Fe_3GeTe_2$, have been explored in a heterostructure with TIs [13-17]. While previous studies have mainly employed a few TI materials, which are tetradymite-type TI compounds such as $(Bi,Sb)_2Te_3$ and $Bi_2Se_3$ as representative TIs [18], expanding the material choices to include alternative TI systems could expand the possibilities for the interface engineering and for exploring additional emergent phenomena.

$\beta$-$Ag_2Te$ has recently been predicted to be a TI based on theoretical calculations [19], and this prediction has been experimentally supported by the observation of quantum oscillations in magnetoresistance and the half-integer quantum Hall effect [20,21]. Owing to its high carrier mobility and small effective mass [16], $\beta$-$Ag_2Te$ exhibits surface dominant electric transport that is less affected by phonon scattering from bulk states and impurity-induced disorder scattering. Furthermore, because $\beta$-$Ag_2Te$ hosts a low carrier system, it does not require additional chemical doping to tune the Fermi level near the Dirac point. Such electrical transport characteristics properties allow the observation of clean surface Dirac states with minimal bulk contributions, making $\beta$-Ag$_2$Te a promising material platform for exploring emergent phenomena in combination with FMIs/FMMs. However, previous studies on $\beta$-Ag$_2$Te have mainly focused on small bulk samples such as nanoribbons or nanoplates [20-25], which are not suitable for fabricating heterostructures. The difficulties in growing high-quality thin films likely originate from the structural mismatch between monoclinic $\beta$-Ag$_2$Te and conventional substrates. The development of high-quality epitaxial $\beta$-Ag$_2$Te thin films is essential for the systematic exploration of these emergent phenomena.

In this study, we report the epitaxial growth of $\beta$-Ag$_2$Te thin films by molecular beam epitaxy. The thin films were fabricated by depositing Ag on InP substrate at room temperature, followed by annealing in the presence of Te. X-ray diffraction (XRD) and transmission electron microscopy (TEM) analysis confirmed that the films grew epitaxially along (002) orientation. Temperature dependent measurement of longitudinal and Hall resistances revealed that two-dimensional conduction is dominant at low temperatures, where three-dimensional bulk carriers deactivate. The successful growth of $\beta$-Ag$_2$Te in thin-film form not only expands the diversity of candidate topological insulator materials beyond the conventional Bi- and Sb-based families but also paves the way for device development, particularly the design of heterojunctions that were previously inaccessible with nanoribbon or nanoplate samples.

The crystal lattice of $\beta$-Ag$_2$Te adopts a distorted antifluorite structure (Fig. 1(a)), with a monoclinic symmetry (space group $P2_1/c$). This phase undergoes a structural phase transition above 140 °C to $\alpha$-Ag$_2$Te, which has a cubic structure [26]. In bulk crystals, the (002) plane of $\beta$-Ag$_2$Te is one of the thermodynamically stable planes [23]; it originates from the (111) plane of $\alpha$-Ag$_2$Te and exhibits a slightly distorted triangular geometry. On the basis of this structural relationship, we selected InP(111)A as the substrate for epitaxial growth, as it also provides a triangular surface lattice. The film growth followed the procedure schematically illustrated in Fig. 1(b). Prior to the thin film growth, the substrates were annealed at 400 °C in a vacuum chamber with a base pressure of approximately $1.0 \times 10^{-7}$ Pa. After cooling to room temperature (~ 30 °C), Ag was deposited under an equivalent beam pressure of $P_{Ag} = 1.0 \times 10^{-5}$ Pa for 30 minutes. Subsequently, the samples were annealed to 250 °C under a Te flux for 30 minutes. Then, the film was cooled down to room temperature in vacuum without additional supply of Ag or Te.

We examined the dependence of film quality on the amount of supplied Te flux during annealing process. Three samples were prepared with different Te fluxes $P_{Te} = 1.0 \times 10^{-5}$, $5 \times 10^{-5}$ and $2 \times 10^{-4}$ Pa, which we refer to as samples A, B, and C, respectively. Figures 2(a)-(c) display the XRD $\theta$-$2\theta$ scans for these samples, where diffraction peaks from (002), (004) and (006) planes of $\beta$-Ag$_2$Te are observed at $2\theta$ = 24.0°, 49.1° and 77.2°, respectively. The spacing between (002) planes was calculated to be 3.71 Å for all samples, in good agreement with the reported value [27]. In sample A (Fig. 2(a)), which was grown with the lowest Te flux, an additional impurity peak from Ag (111) is observed at $2\theta = 37.8°$ (represented as a lower triangle), whereas no such extra peak appears in samples B and C. Sample B exhibits pronounced Laue fringes (Fig. 2(d)), indicating coherent ordering of lattice planes throughout the entire film. In addition, the rocking curve around the (002) reflection of sample B shows a narrow full width at half maximum (FWHM) of 0.10° (Fig. 2(e)), in sharp contrast to sample C, whose FWHM is 0.40° (Fig. 2(e)). These results highlight that the amount of supplied Te is a key parameter to control the structural quality of epitaxial $\beta$-Ag$_2$Te thin films. We note that the film thickness of sample B is calculated to be 29 nm from X-ray reflectometry.

Figure 3(a) shows a high-angle annular dark-field (HAADF)-scanning transmission electron microscopy (STEM) image the $\beta$-Ag$_2$Te thin film (Sample B). The clear atomic-resolution contrast indicates the highly aligned $\beta$-Ag$_2$Te crystal lattice along the lateral direction, demonstrating that the $\beta$-Ag$_2$Te film obtained by the present growth procedure is free from major structural defects. The film thickness measured from the STEM image is approximately 30 nm, which is consistent with the value estimated form X-ray reflectometry. We performed a fast Fourier transform (FFT) analysis on the film area that is indicated as red square in Fig. 3(a). The obtained FFT pattern (Fig. 3(c)) shows good agreement with the calculated reciprocal pattern of (002)-oriented $\beta$-$Ag_2Te$ thin film (Fig. 3(d)) in terms of both the positions and relative intensities of the diffraction spots. Figures 3(e)-(h) respectively display the EDX mappings for In, P, Ag and Te, while Fig. 3(b) shows the corresponding atomic concentration depth profiles. These results confirm that each element is uniformly distributed throughout the main body of the film, with no apparent elemental gradient along the thickness direction. We observed a non-uniform region near the film-substrate interface (approximately 2–3 nm thick), which we attribute to an interfacial oxide layer or a sacrifice layer caused by the lattice mismatch between the cubic InP substrate and the monoclinic $\beta$-$Ag_2Te$. However, this structural disorder is confined to the interface, and no additional crystalline phase or elemental segregation was observed at the vacuum surface. Apart from this interfacial region, these results demonstrate the successful growth of a chemically homogeneous, single-crystalline $\beta$-$Ag_2Te$ film aligned with the InP substrate.

We examined the electrical transport properties for sample B. The measurements were performed using a standard four-probe method on 2 mm × 5 mm-sized sample. Longitudinal and Hall resistances were measured up to 7 T with a Physical Property Measurement System (Quantum Design). In the following discussion, we present sheet longitudinal and Hall resistivities, denoted as $R_{xx}$ and $R_{yx}$, respectively, where the aspect ratio of the sample is taken into account. Figure 4(a) displays the temperature dependence of $R_{xx}$. As the temperature decreases from 300 K, $R_{xx}$ increases steeply, then decreases below 60 K, and rises again below 10 K. This nonmonotonic temperature dependence resembles that observed in bulk $\beta$-$Ag_2Te$ samples [20,21], and is also commonly observed in carrier-controlled TIs such as $(Bi,Sb)_2Te_3$ and $Bi_2(Te,Se)_3$ [28,29]. In those systems, the high-temperature insulating behavior and the low-temperature metallic behavior are attributed to bulk and surface conduction, respectively. To qualitatively evaluate the contribution of the two conduction channels in our sample, we performed a fitting analysis based on a parallel conduction model consisting of three-dimensional semiconducting bulk states and two-dimensional metallic surface or interface states, which is previously applied to bulk $\beta$-$Ag_2Te$ samples [21,30]. The fitting equation for the temperature dependence of $R_{xx}$ is given by,

$$\frac{1}{R_{xx}} = \frac{1}{(AT + R_s)} + \frac{1}{R_b}\exp\left(-\frac{E_a}{k_B T}\right) \quad (1)$$

, where $R_s$ and $R_b$ represent the resistivity of the bulk state and the low-temperature saturating resistivity of the two-dimensional metallic state, respectively. $A$ accounts for the temperature dependence of resistivity for the two-dimensional state, and $E_a$ denotes the activation energy from the Fermi level $E_F$ to bulk band edge. As shown by the red curve in Fig. 4(a), this fitting reproduces the experimental results well, yielding the parameters $R_s$ = 11 k$\Omega$, $R_b$ = 0.52 k$\Omega$, $A$ = 0.15 k$\Omega$/K and $E_a$ = 30 meV. The extracted $E_a$ value is consistent with the previous reports on bulk samples (29 ~ 33 meV), despite $R_{xx}$ in bulk being two orders of magnitude smaller than that in our thin films [21]. This consistency of $E_a$ indicates that the bulk band structure and position of $E_F$ are reliably reproduced in thin-film form. Moreover, the bulk band gap has been reported to be 64 ~ 80 meV from optical measurements [31,32] and ~ 80 meV from DFT band calculations [19], which is consistent with the present value, assuming that $2E_a$ corresponds to the bulk band gap.

Figure 4(b) presents sheet Hall resistivity $R_{yx}$ as a function of magnetic field $B$ at various temperatures. To eliminate the contribution from longitudinal resistance, the raw data of Hall resistance were antisymmetrized with respect to $B$. At low temperatures ($T$ < 30 K), $R_{yx}$ increases linearly with $B$, indicating that p-type (hole) conduction dominates. In contrast, at high temperatures ($T$ > 200 K), $R_{yx}$ is also linear in $B$ but with a negative slope, indicating n-type (electron) transport. From the discussion of the temperature dependence of $R_{xx}$, the hole conduction at low temperature is consistent with a two-dimensional conduction channel, while the electron conduction at high temperature can be attributed to a three-dimensional bulk channel. This interpretation is further supported by the $B$-nonlinear behavior of $R_{yx}$ in the intermediate temperature region, reflecting the coexistence of both electrons and holes. Figures 4(c) and (d) show the temperature dependences of $1/eR_H$ and electron and hole mobilities $\mu_e$ and $\mu_h$, respectively. $1/eR_H$ corresponds to the sheet carrier density, whose sign indicates the carrier type (positive for hole density $p$ and negative for electron density $n$). Here, $R_H$ and $e$ denote the Hall coefficient and the elementary charge, respectively. The carrier densities were estimated from the slope of $R_{yx}$ against $B$. The gray shaded region corresponds to the temperature range where multi-carrier transport is observed. At high temperatures, the electrons become thermally activated and gradually diminish with decreasing temperature, showing semiconducting behavior. In contrast, at low temperatures, the hole density remains nearly constant with temperature, consistent with metallic transport, with $p = 1.0 \times 10^{12}$ cm$^{-2}$ and $\mu_h = 400$ cm$^2$V$^{-1}$s$^{-1}$, respectively, at $T = 2$ K. We prepared a series of samples with varying thicknesses, grown with similar conditions as sample B, to examine their electrical transport properties. The thickness dependence of the carrier density and conductivity further supports a model in which the low-temperature conduction channel is confined to the interface or surface (See supplementary materials S1 and S2). One possible origin of this two-dimensional conduction is the topological surface states of $\beta$-Ag$_2$Te. However, we note that the non-uniform interfacial region at the Ag$_2$Te-InP boundary could also serve as a two-dimensional conduction channel through an oxidized or sacrificial layer. Further structural and spectroscopic investigations will be needed to clarify the origin of the two-dimensional conduction.

If the two-dimensional conduction originates from topological surface states with the degenerated top and bottom surface states, the corresponding Fermi energy is estimated to be approximately 84 meV below the Dirac point from the relationship of two-dimensional Dirac states, $E_F = \hbar v_F \sqrt{2\pi p}$. Here, we used a Fermi velocity of $5.0 \times 10^5$ m/s for the surface Dirac dispersion [22, 24]. Because the estimated Fermi energy is comparable to the bulk band gap and the Dirac point is located within the band gap according to the band calculation [19], we consider that electron potential becomes higher near the surface of the film (see the schematic band diagram in the inset of Fig. 4(a)), which can explain the bulk electron transport at elevated temperature and surface hole transport at low temperature. We note that bulk samples predominantly showed n-type surface carriers, whereas the present thin film samples exhibit p-type surface conduction.

In summary, we have demonstrated epitaxial growth of $\beta$-Ag$_2$Te thin film by molecular beam epitaxy. By careful tuning of the supplied Te flux, we obtained high-quality epitaxial films. The temperature and thickness dependences of $R_{xx}$ and $R_{yx}$ reveal the two-dimensional metallic conduction at low temperature where three-dimensional bulk carriers are deactivated, which is consistent with previous studies on bulk $\beta$-Ag$_2$Te samples. Whether the two-dimensional conduction originates from topological surface states or from the interfacial region near the Ag$_2$Te-InP boundary remains an open question, and further measurements will be needed to clarify the origin. Nevertheless, the epitaxial $\beta$-Ag$_2$Te thin films obtained in this work enable the fabrication of heterojunctions and overcome the limitations of previous bulk and nanostructure studies, offering a scalable route for developing advanced topological devices.

**FIGURES**

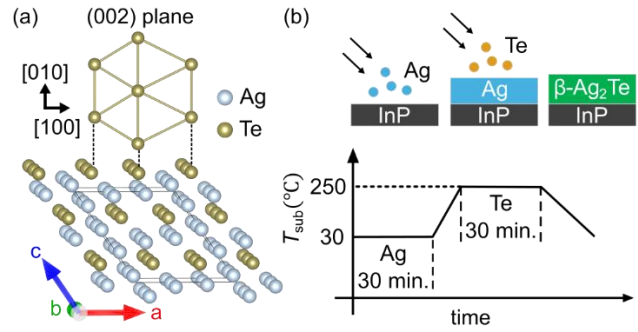

Figure 1 (a) Crystal structure of β-Ag$_2$Te ($a$ = 8.162 Å, $b$ = 4.467 Å, $c$ = 8.973 Å) and (002) lattice plane. (b) A schematic diagram for the growth procedure of β-Ag$_2$Te thin film.

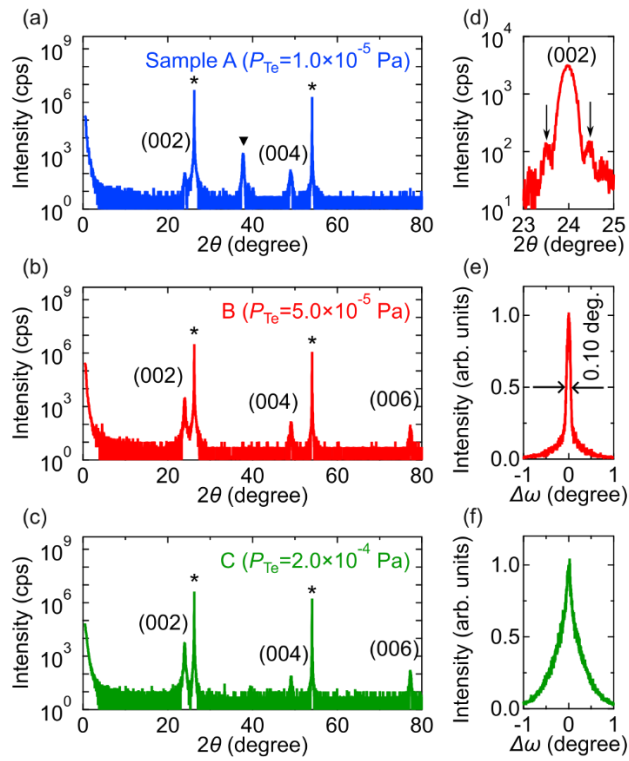

Figure 2 (a-c) $\theta$-$2\theta$ scans of X-ray diffraction for samples A-C with different Te flux. (111) and (222) diffraction peaks of InP substrate are represented with asterisks. The peak denoted by a lower triangle in (a) represents the diffraction from (111) lattice plane of Ag. (d) An enlarged view of the $\theta$-$2\theta$ scan around the (002) $\beta$-Ag$_2$Te peak in (b). (e, f) Rocking scans around the $\beta$-Ag$_2$Te (002) peaks of samples B and C.

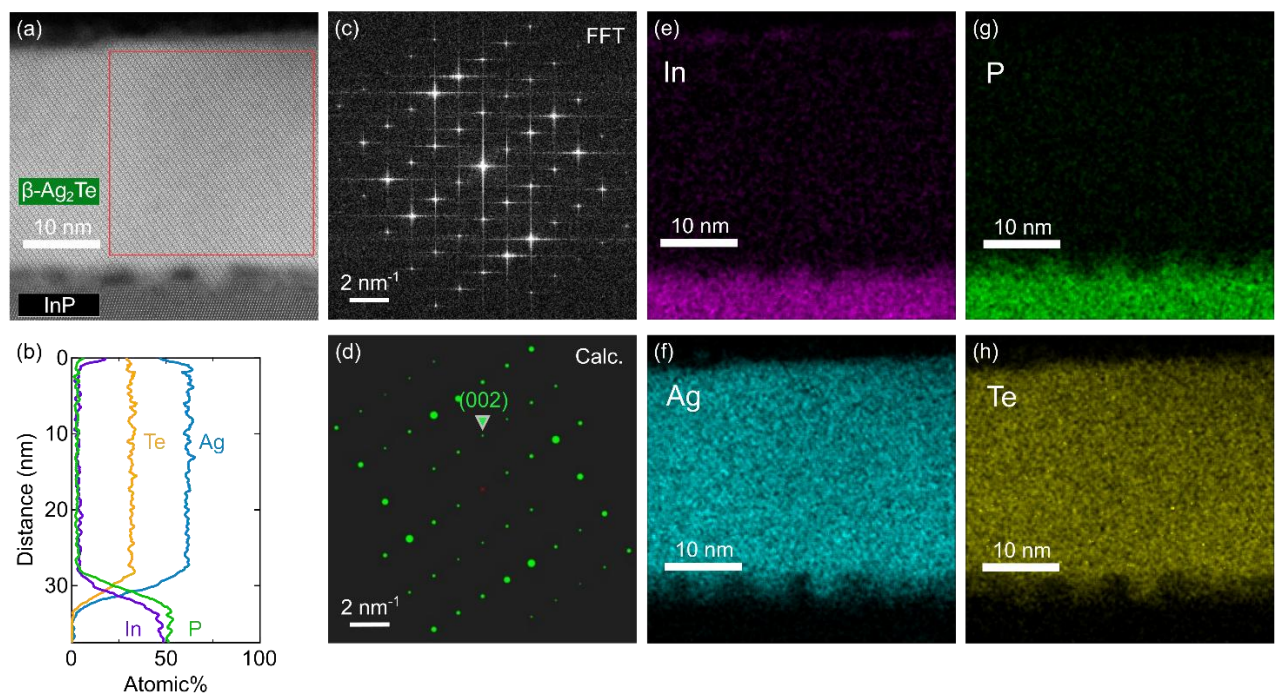

Figure 3 (a) High-angle annular dark-field scanning transmission electron microscopy (HAADF-STEM) image of sample B ($\beta$-Ag$_2$Te thin film). (b) Atomic concentration depth profiles derived from the EDX analysis. (c) Fast Fourier transform (FFT) image obtained from the area indicated by the red square in (a). (d) Calculated reciprocal lattice pattern of $\beta$-Ag$_2$Te for the [0-10] beam incidence. We used ReciPro to calculate the diffraction patterns.[33] (e-h) Energy dispersive X-ray spectroscopy (EDX) elemental mappings of (e) In, (f) Ag, (g) P, and (h) Te.

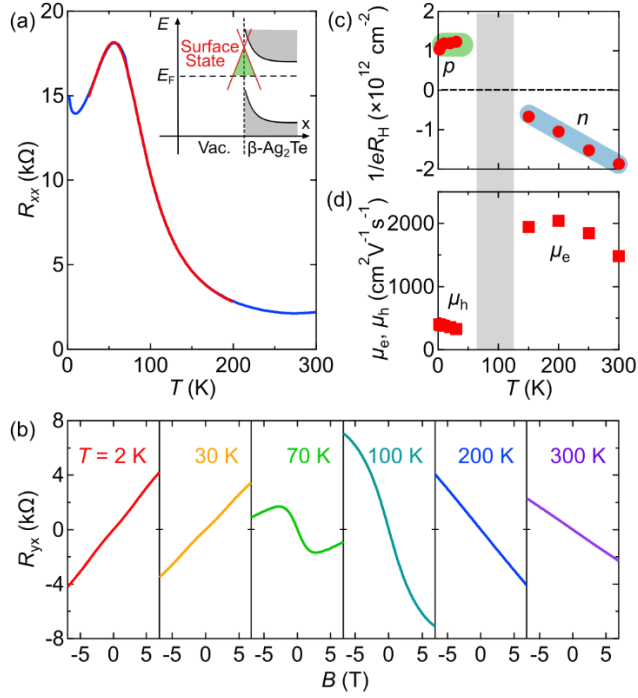

Figure 4 (a) Temperature dependence of longitudinal sheet resistivity $R_{xx}$. The red solid line represents the fitting curve assuming a parallel conduction of surface and bulk. The inset shows a schematic of the band structure near the surface of the film. (b) Magnetic field dependence of sheet Hall resistivity $R_{yx}$ for various temperatures. (c) Temperature dependence of $1/eR_H$, corresponding to the sheet carrier density, where the sign indicates the carrier type (holes or electrons). $R_H$ and $e$ denote the Hall coefficient and elemental charge, respectively. (d) Temperature dependence of electron and hole mobility $\mu_e$ and $\mu_h$. In (c) and (d), the gray shaded region from 70 K to 120 K corresponds to the temperature region where multi carrier conduction is observed in the Hall measurement.

## ACKNOWLEDGEMENTS

We thank D. Hamane for the transmission electron microscope experiments. We are also grateful to M. Urai and R. Takagi for assistance with low temperature transport measurements, and A. Tsukazaki for fruitful discussions on thin-film fabrication. This work was supported by JSPS KAKENHI (Grant No. 23K26554), Toyota Physical and Chemical Research Institute, and Mitsubishi Foundation. This article has been submitted to Applied Physics Letters (AIP Publishing). After it is published, it will be found at https://pubs.aip.org/aip/apl.